\documentclass[aps,prl,twocolumn,superscriptaddress,showpacs,amsfonts,amsmath,amssymb]{revtex4-1}

\usepackage{graphicx}
\usepackage{amsmath, amssymb}
\usepackage{color}

\definecolor{Red}{rgb}{1,0,0}
\def\B{\color{Red}}
\definecolor{Black}{rgb}{0,0,0}
\def\B{\color{Black}}

\begin{document}
\title{A reversible conversion between a skyrmion and a domain-wall pair in junction geometry}
\author{Yan Zhou}
\affiliation{Department of Physics, The University of Hong Kong, Hong Kong, P. R. China}
\affiliation{Center of Theoretical and Computational Physics, Univ. of Hong Kong, P. R. China}
\author{Motohiko Ezawa}
\affiliation{Department of Applied Physics, University of Tokyo, Hongo 7-3-1, 113-8656, Japan}

\begin{abstract}
A skyrmion is a topological texture in the continuum field theory.
Recent experimental observation of skyrmions in chiral magnet evokes a flourish of its extensive study.
Skyrmion is expected to be a key component of the next generation spintronics device called \textquotedblleft skyrmionics\textquotedblright.
On the other hand, there is a well established memory device encoded by a sequence of domain walls.
A skyrmion carries a topological number, whereas a domain wall does not.
Nevertheless, we show a conversion is possible between a skyrmion and a domain-wall pair by connecting wide and narrow nanowires, enabling the information transmission between skyrmion device and domain-wall device.
Our results will be the basis of a hybrid device made of skyrmions and domain walls,
where the encoded information in domain walls is converted into skyrmions, and then read out by converting the skyrmions back to domain walls
after a functional control of the skyrmions.
\end{abstract}
\maketitle

A skyrmion is a spin texture with a quantized topological number.
Recently, skyrmions have been observed experimentally in chiral ferromagnets without inversion symmetry\cite{Mohlbauer,Munzer,Yu2010,Pfleiderer,Yu2011}.
It is stabilized dynamically by the Dzyaloshinskii-Moriya interaction (DMI) and external magnetic field\cite{Bogdanov1989,Bogdanov1994,Rossler,Han,Ezawa2011}. Once it is created, its stability is topologically protected
against dissipation and fluctuation even in the presence of large defects \cite{Sampaio,Fert}.
The position of a skyrmion is controllable by applying current.
A train of skyrmions in magnetic perpendicular magnetic anisotropy (PMA) nanowires with DMI {\B has} been proposed to be a novel alternative as an information carrier\cite{Sampaio,Fert}, which is called  \textquotedblleft skyrmionics\textquotedblright .
However, the generation of an isolated skyrmion remains a challenging task.
Indeed, it is impossible to change continuously the topological number from zero to its quantized value
and to create a topological object without breaking the continuity of the field by such as applying a laser beam\cite{Marco,Ezawa2010} or applying a circulating current\cite{Tchoe}.
This is actually the case only in a sufficiently large system.
A skyrmion can also be created or destroyed when a skyrmion touches an edge since the spin can rotate at the edge. A recent theoretical study has shown that a skyrmion can be created from a notch of a nanowire by applying an electrical current\cite{IwasakiNnano}.

On the other hand, current-induced domain wall (DW) motions in magnetic thin film have been extensively studied in recent years towards high-efficiency and low-dissipation spintronic memory. Many novel concepts and designs based on current-driven DW motions have been proposed theoretically and realized experimentally including the racetrack memory,  where a train of DWs are moved along a nanowire, for ultradense information storage and information processing\cite{Parkin,Hayashi}. Domain wall in ultrathin magnetic  films  layered between a heavy  metal and an oxide  with  PMA  \cite{Miron1,Miron2,Haazen,LiuluqiaoPRL} shows anomalously large velocity under current induced spin transfer torque (STT). Such intriguing results generated competing explanations of their underlying physical mechanisms \cite{Miron1,Miron2,Liuluqiao,Brataas2014}. More recently, chiral DWs with a preferred handedness that is influenced by DMI have been observed which  triggers intensive research efforts for emerging spintronics applications based on  chiral magnetism\cite{Thiaville,Su,Emori,Chen2013}.
While DW spintronics has attracted tremendous interest for the potential applications of ultrahigh-density information processing devices, the central issues are further reduction of  the power consumption and the  joule heating.
By contrast, the critical current density can be very low to drive a skyrmion\cite{IwasakiNcom}.

It is intriguing that both skyrmions and DWs are accommodated albeit different geometries
in non-centrosymmetric bulk materials or thin film heterostructure lacking inversion symmetry.
Here, we focus on the following issues on a DW pair and a skyrmion, though a skyrmion has a topological number whereas a DW pair does not:
(a) Is it possible to realize the integration of DW and skyrmion devices,
such that they can be mutually converted to each other in one system?
(b) Is it possible to combine important properties and advantages of these two very different magnetization textures?
Our answers are affirmative. We present an explicit conversion mechanism between a skyrmion and a DW pair by employing nanowires with different widths.
We propose a hybrid functional device where the information is encoded as a train of DWs and then converted into a train of skyrmions. After a functional control of the skyrmions, the information is read out by converting the skyrmions back to the DWs.

\begin{figure*}
\begin{center}
\includegraphics[width=0.9\textwidth]{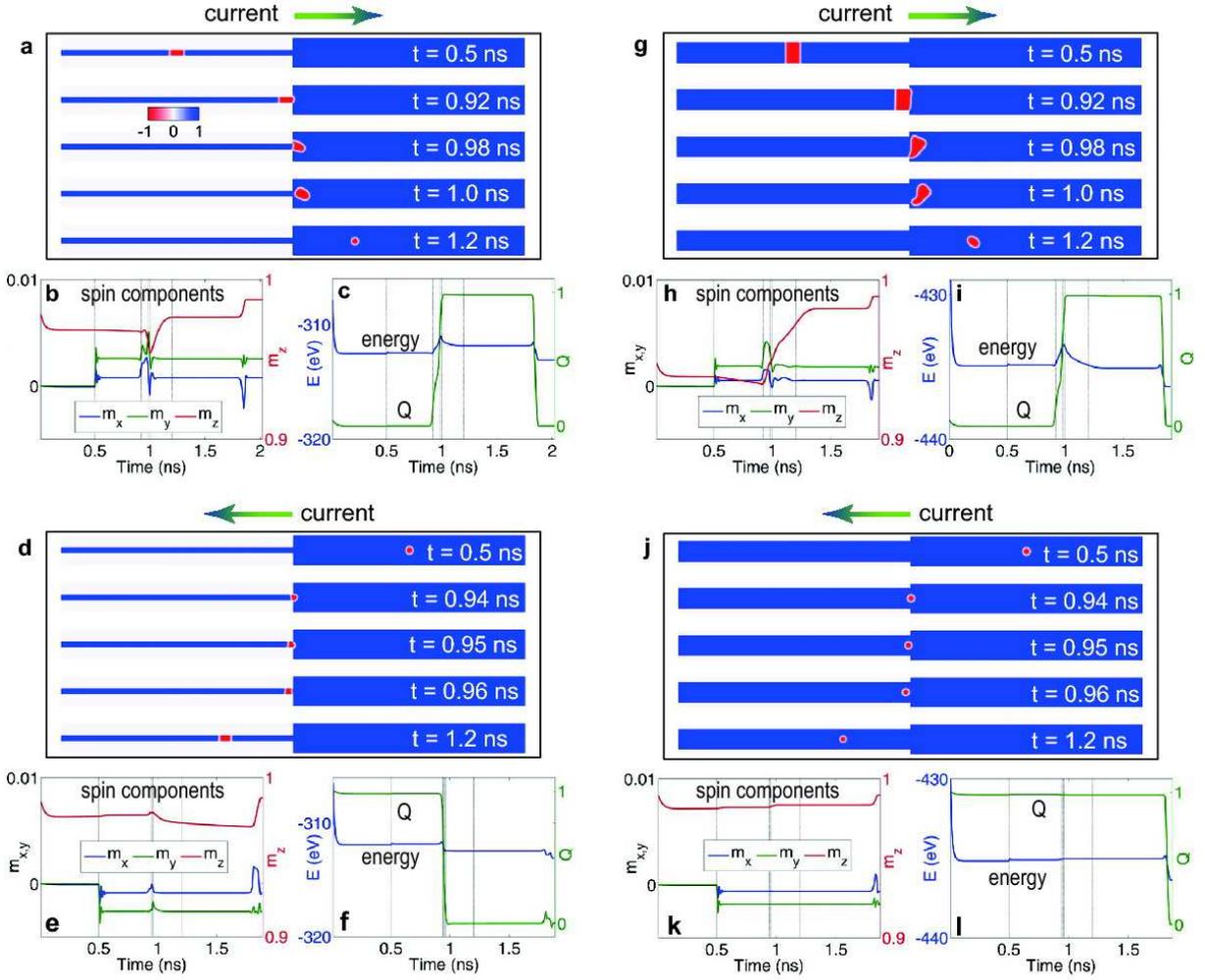}
\end{center}
\caption{\textbf{Conversion between a skyrmion and a DW pair.} The in-plane current density (along $\pm x$ direction) is 8 $\mu$A/nm$^2$.
\textbf{a-c}:  A DW pair is converted into a skyrmion  driven by spin transfer torque from the narrow part to the wider part of the nanowire.
\textbf{a} The snapshots of the magnetization configuration at five selected times corresponding to the five vertical lines in \textbf{b} and \textbf{c}.
\textbf{b} The time evolution of the total spin component $m_x,m_y,m_z$.
\textbf{c} The time evolution of the energy and the Pontryagin number $Q$.
\textbf{d}$-$\textbf{f} Reverse process of \textbf{a}$-$\textbf{c}.
\textbf{g}$-$\textbf{i}  Conversion process of a DW pair into a skyrmion similar to \textbf{a}$-$\textbf{c},
when the width of the narrow part is larger than the skyrmion intrinsic size.
\textbf{j}$-$\textbf{l} Reverse process of \textbf{g}$-$\textbf{i}, where a skyrmion is not converted into a DW pair.
}
\label{fig:mutualconversion}
\end{figure*}

\section*{Skyrmion and Topological Stability}
A skyrmion is a texture of the normalized classical spin $\boldsymbol{n}=(n_x,n_y,n_z)$
which has a quantized topological number.
Under an external magnetic field, the ground-state spin is up, \emph{i.e.}, $(0,0,1)$,
while the spin direction of the skyrmion center is down, \emph{i.e.}, $(0,0,-1)$.
Spins swirl continuously around the core and approach the ground-state value asymptotically.
The skyrmion is characterized by the Pontryagin number $Q$, which is the topological number in the planar system, $Q=\int d^2x\rho _{\text{sky}}(\boldsymbol{x})$, with
\begin{equation}
\rho _{\text{sky}}(\boldsymbol{x})=-{\frac{1}{4\pi }}
\boldsymbol{n}(\boldsymbol{x})\cdot \left( \partial _{x}\boldsymbol{n}(\boldsymbol{x})\times
\partial _{y}\boldsymbol{n}(\boldsymbol{x})\right).  \label{PontrNumbe}
\end{equation}
The topological number density $\rho _{\text{sky}}(\boldsymbol{x})$ may be expressed as a total derivative, and hence it depends only on the boundary value. We obtain $Q=1$ for a skyrmion in a sufficiently large area.
Consequently, even if the skyrmion spin texture is deformed, its topological number does not change, as far as the boundary condition is not modified. It can be neither destroyed nor separated into pieces, \emph{i.e.}, a skyrmion is topologically protected.

The necessary conditions of the topological conservation are the continuity of the spin texture and the boundary condition imposed on it. On one hand, the continuity of the spin texture is destroyed for instance by applying a laser or considering a microscopic phenomenon comparable with the lattice size.
On the other hand, when a skyrmion touches an edge of the sample, the topological number may continuously change. Then we can create or destroy a skyrmion.

The main topic of this work is the demonstration of mutual conversion between a skyrmion which is a topological object ($Q=1$) and a DW pair which is non-topological object ($Q=0$).
Once the skyrmion is created and moves into a sufficiently large sample without touching the edge, it is a stable object  due to its topological protection.

There exist two types of  DMI, depending on samples\cite{Fert,Bogdanov1989,Bogdanov1994}. It determines how spins cant at the edge of a sample and how spins swirl around the skyrmion core. In this work we assume the Neel-type (hedgehog) skyrmion, although essentially the same conclusions follow when we start with the Bloch-type (vortex) skyrmion.

\section*{Creation of skyrmion from domain wall and vice versa}
Chiral DWs are expected for a magnetic PMA thin film on a substrate inducing DMI. Chiral structures such as skyrmions are also stabilized in such a system. We consider a nanowire geometry with the width $W$.
A skyrmion has an  intrinsic radius $R$ = 28 nm  determined by the material parameters and nanowire geometry given in METHODS.
A skyrmion is stable for $W> 2R$, while it cannot exist for $W<2R$, where the resultant structure is found to be a DW pair.

\begin{figure*}
\begin{center}
\includegraphics[width=0.8\textwidth]{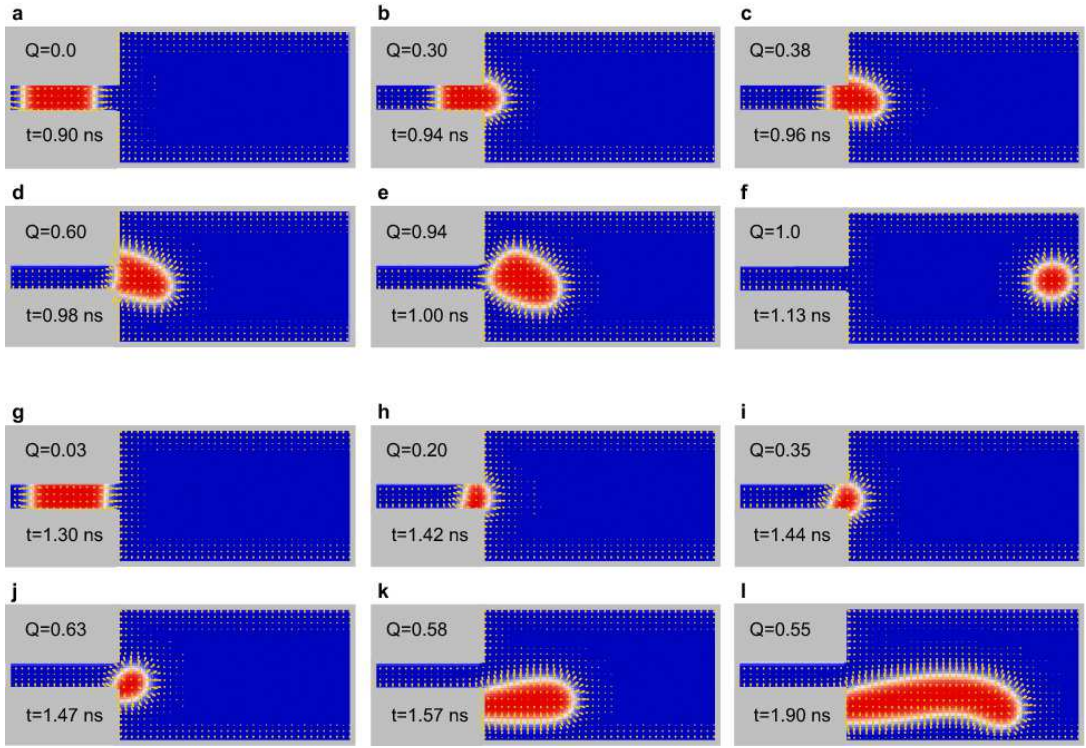}
\end{center}
\caption{\textbf{Spin configurations of a skyrmion, a meron and a DW pair.}
\textbf{a}$-$\textbf{f}  A skyrmion is created from a DW pair driven by spin transfer torque from the narrow part to the wide part of the nanowire. The spin direction is shown by yellow arrow.
A yellow dot in the blue region shows that the spin is almost up,
while a yellow dot in the red region shows that it is almost down.
No dots are observed when the spin is almost perfectly up or down.
The in-plane current density (along $\pm x$ direction) is 8 $\mu$A/nm$^2$.
\textbf{g}$-$\textbf{l} A meron is created when the current is 5 $\mu$A/nm$^2$.}
\label{fig:creation}
\end{figure*}

We investigate a junction comprised of a narrow nanowire ($W<2R$) on the left side and a wide nanowire ($W>2R$) on the right side: See Fig. \ref{fig:mutualconversion}a.
We apply a current parallel to the nanowire. According to the Thiele equation, a spin texture moves with the same velocity as the current without changing its shape\cite{Thiele}.
By applying a current from the left to the right, a DW pair moves in the same direction and then is converted into a skyrmion (Supplementary Movie S1).

Fig. \ref{fig:mutualconversion}b shows a time-evolution of the total spin
$\boldsymbol{m}=(m_x,m_y,m_z)$.
$m_x$ and $m_y$ is almost zero between $0<t<0.5$ ns during the relaxation process.
At $t=0.5$ns, the finite in-plane spin emerges, which is due to the STT from the injected current.
The time-evolution of the energy is shown as the blue curve in Fig. \ref{fig:mutualconversion}c.
The energy of the DW pair is slightly lower than that of a skyrmion.
Fig. \ref{fig:mutualconversion}c shows the Pontryagin number changes continuously between one for a skyrmion and zero for a DW pair (see the green curve).
The Pontryagin number can change its value when it interferes with an edge.
The main message is that a skyrmion and a DW pair can be deformed continuously.
It should be noted that we have extracted the background topological number $-0.06$, which is caused by the spin canting at the edges. A skyrmion is destroyed when it touches the sample edge. Accordingly, the Pontryagin number becomes zero after the skyrmion is expelled from the right edge of the sample.

Next we investigate the reverse process.
We start with a skyrmion in the wider part of the nanowire (i.e., the right half of the entire nanowire): See Fig. \ref{fig:mutualconversion}d.
We apply a current from right to left.
The skyrmion moves from right to left and is converted to a DW pair since it can not exist in the narrow nanowire whose width is less than $2R$.
This result shows that a conversion between a skyrmion and DW pair is a reversible process (Supplementary Movie S2).

We proceed to study the case where we start with a DW pair in the left part of the nanowire whose width is larger than $2R$ in Fig. \ref{fig:mutualconversion}g.
A skyrmion is created in a similar way as in Fig. \ref{fig:mutualconversion}a, except that the created skyrmion in the junction is greatly deformed.
It takes much more time than the previous case (Fig. \ref{fig:mutualconversion}a)
before the skyrmion changes its shape to its intrinsic size (Supplementary Movie S3).
The (absolute value) of total energy for this case is larger than Fig. \ref{fig:mutualconversion}a since the total energy is proportional to the width of the nanowire.

An interesting situation occurs when we consider the reverse process [Fig. \ref{fig:mutualconversion}j].
A skyrmion is not converted into a DW pair but remains to be a skyrmion in the left part of the nanowire.
A conversion does not occur since the width of the left part of the nanowire is larger than $2R$.

Let us describe the conversion process from a DW pair to a skyrmion in more details.
First we start with a DW pair in the narrow part of the nanowire: See Fig. \ref{fig:creation}a.
As an electrical current is applied from the left to the right, the DW pair also moves in the same direction of the applied current without changing its shape.
When the right DW reaches the junction, both the endpoints of the DW are pinned at the junction whereas the central region of DW continues to move.
As a result, the DW is deformed into a curved shape: See Fig. \ref{fig:creation}b.
Next, the anti-DW also reaches the junction,
and a skyrmion-like structure is formed. This structure still touches the edge at t = 0.98 ns (Fig. \ref{fig:creation}d), and then it is decoupled from the edge at t = 1 ns (Fig. \ref{fig:creation}e). The topological number is not exactly one due to the interference between the skyrmion halo and the edge (Fig. \ref{fig:creation}e). Finally the skyrmion size shrinks to the intrinsic radius, and the resultant structure becomes a perfect circular shape with $Q=1$ (Fig. \ref{fig:creation}f).

\section*{Meron generation}
\begin{figure*}
\begin{center}
\includegraphics[width=0.8\textwidth]{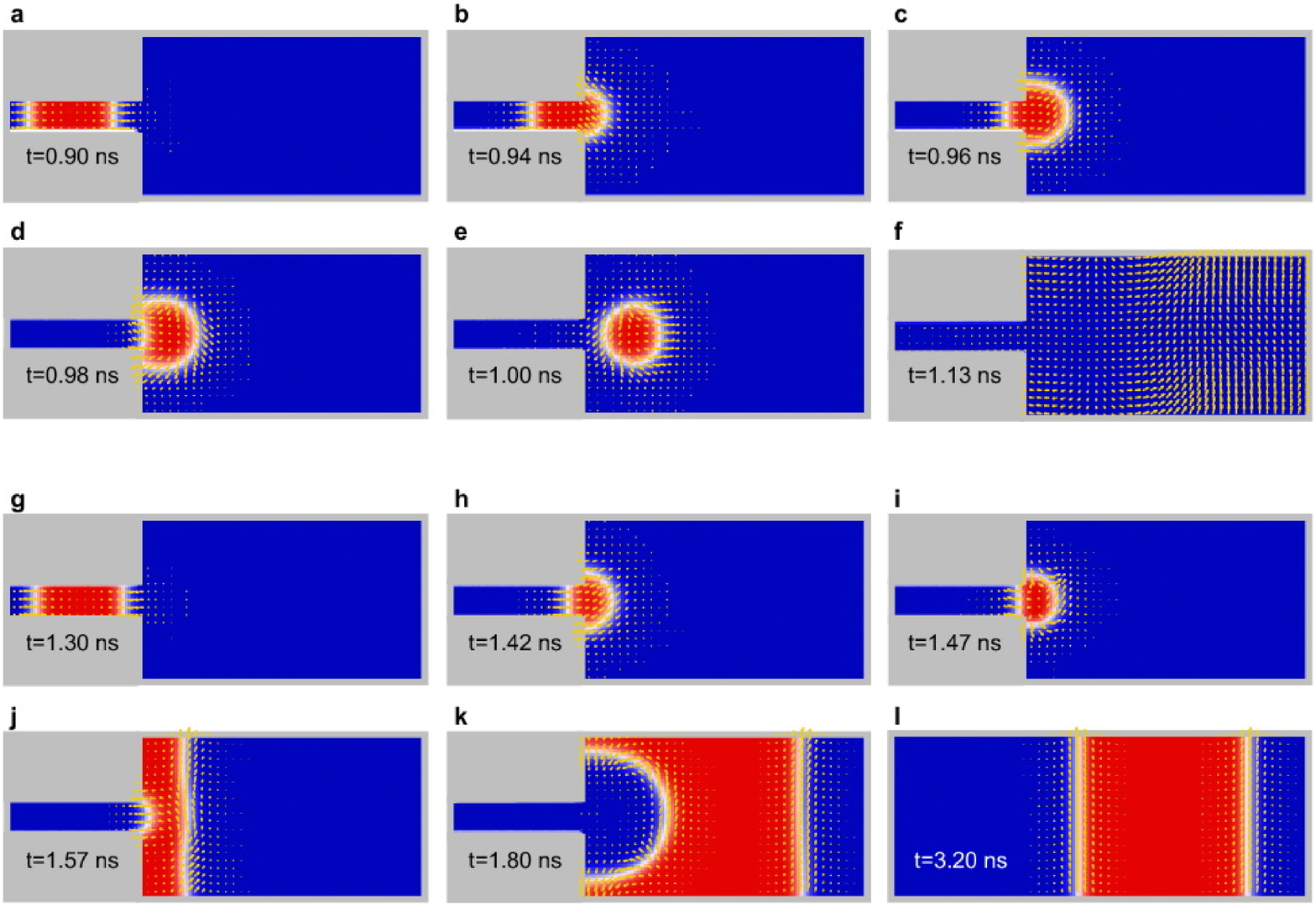}
\end{center}
\caption{\textbf{Magnetic droplet soliton without DMI.}
A similar process to Fig. \ref{fig:creation} but without the DMI.
\textbf{a}$-$\textbf{f}
A magnetic droplet soliton is formed from a DW pair, and then disappears {\B by emitting spin waves}.
It does not have the topological number ($Q=0$). The in-plane current density (along $\pm x$ direction) is 8 $\mu$A/nm$^2$.
\textbf{g}$-$\textbf{l} A DW pair remains as a DW pair when the current is as low as 5 $\mu$A/nm$^2$.
}
\label{withoutDMI}
\end{figure*}

If the applied current density is reduced,  the result is dramatically changed as shown in Fig. \ref{fig:creation}g$-$l.
For smaller current density, a skyrmion is not formed. The calculated critical current for skyrmion generation is 7.5 $\mu$A/nm$^2$, below which the spin texture is attached to one edge of the nanowire and is elongated (Fig. \ref{fig:creation}l, and Supplementary Movie S4). In this case, the applied current density is  5 $\mu\text{A/nm}^2$.
The topological number is almost one half, implying that the structure is a meron\cite{Ezawa2011} which is a half-skyrmion.

\section*{String geometry}
It is intriguing to recapitulate the conversion mechanism between a DW pair and a skyrmion in terms of a string geometry. The bulk is separated into two regions with $n_z>0$ (blue) and $n_z<0$ (red) by the boundary with $n_z=0$ (white): See Fig. \ref{fig:creation}.
The spin direction is almost perpendicular at the boundary due to the DMI effect.
The boundary may be referred to as a string.

A string is either closed or open. It has no branches. An open string must touch the edge.
A DW pair consists of two parallel open strings, where the spin directions imply that $Q=0$.
We may argue that there are only two possibilities how these two open strings develop.

In the first case,
one open string reaches the edge and is expelled [Fig. \ref{fig:creation}b]. Then it makes a semicircle. When the other open string reaches the interface separating the narrow part and the wide part of the nanowire, two open strings can form one closed string. The spin direction of a closed string must form a hedgehog with $Q=1$. It develops into a skyrmion.

In the second case,
when the velocity is small, an open string may cant near the interface [Fig. \ref{fig:creation}h].
The upper part of the DW breaks away from the narrow part of the nanowire while the bulk of the spins is attached to the lower endpoint [Fig. \ref{fig:creation}i], leading to an open string in the end [Fig. \ref{fig:creation}j].
It develops into a meron.

\section*{Non-topological magnetic solitons without DMI}
The DMI plays two roles.
First, the Hamiltonian without the DMI is invariant under the reflection $y \mapsto -y$.
Hence the asymmetric behavior of a skyrmion in Fig. \ref{fig:creation} is due to the DMI.
The spin configuration forming a hedgehog is also due to the DMI.
Second, a skyrmion is stabilized dynamically by the DMI.
Although the topological number $Q$ is defined even without the DMI, it does not play any role if a hedgehog spin configuration is not formed to generate $Q=1$.

\begin{figure*}
\begin{center}
\includegraphics[width=0.48\textwidth]{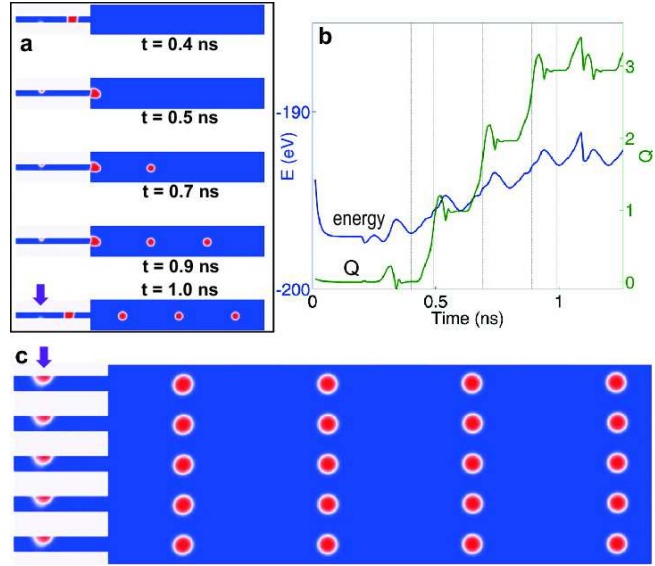}
\end{center}
\caption{\textbf{Skyrmion racetrack}
\textbf{a} Top-view of the spin structure at five different times of the simulation, corresponding to the five vertical lines in \textbf{b}. \textbf{b} Time-evolution of the total energy and the topological number $Q$. \textbf{c} Skyrmion array generations. $Q$ increases by one unit as one skyrmion is created.
The purple arrow {\B represents} the position where an external AC magnetic field is applied to generate DW pairs at a constant time interval.}
\label{fig:racetrack}
\end{figure*}

In order to clarify the important roles of the DMI, we have simulated the above process without the DMI.
Fig. \ref{withoutDMI}a$-$f show a similar process to Fig. \ref{fig:creation}a$-$f in the absence of the DMI.
The magnetization configuration has zero Pontryagin number, which implies that the resultant structure is a non-topological magnetic droplet soliton, which is a novel dynamical magnetic object discovered very recently \cite{Hoefer,Akerman}.
Due to the damping, the magnetic droplet soliton decays after it breaks away from the interface since
it is neither stabilized dynamically nor topologically (Supplementary Movie S5). Spin waves are emitted when the droplet is damped out, as is shown in Fig. \ref{withoutDMI}f.

Fig. \ref{withoutDMI}g$-$l show a similar process to Fig. \ref{fig:creation}g$-$l in the absence of the DMI when the current density is small.
In this case, a DW pair remains as a DW pair but with larger width after some perturbative modulations (see Fig. \ref{fig:creation}l and Supplementary Movie S6).

\section*{Skyrmion racetrack}
In the above, the initial state of the simulations all start with a given spin configuration \emph{i.e.}, either a DW pair or a skyrmion. For practical applications, it is desirable to create a skyrmion from a ferromagnetic ground state.
 We apply a periodic modulation of magnetic field, {\B $B=B_0\sin \omega t $}, close to the left edge of the nanowire. The AC field amplitude is $B_0=1$ T with frequency of $\omega=5$ GHz. A sequence of DWs are created at {\B a regular interval} and then deformed into skyrmions. The micromagnetics simulation results are shown in Fig. \ref{fig:racetrack} (Supplementary Movie S7). The total Pontryagin number takes large value corresponding to the creation of skyrmions. This concept can be extended to a skyrmion array if we make multiple nanowires in parallel and then merge into an extended thin film with larger width as shown in Figure \ref{fig:racetrack}b (Supplementary Movie S8).

\begin{figure*}
\begin{center}
\includegraphics[width=0.48\textwidth]{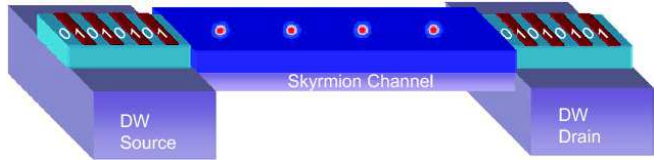}
\end{center}
\caption{\textbf{A hybrid device based on DWs and skyrmions.}
A sequence of information (\emph{i.e.}, data bits) is encoded as a train of DWs and then converted into a train of skyrmions,
which is transmitted in the skyrmion channel. We read out the information by converting skyrmions back to DWs.
}
\label{fig:SharpJunction}
\end{figure*}

\section*{Perspectives}
We have presented a conversion mechanism between a DW pair and a skyrmion [Fig. 2a$-$f].
We have shown that there exists a critical current density below which a DW pair is converted into a meron [Fig. 2g$-$l] rather than a skyrmion. Without DMI, the non-topological droplet-soliton [Fig. 3a$-$f] will be generated in the same system. These intriguing results suggest that the proposed nanowire system may provide an excellent environment for studying a wide range of exotic nano-magnetic phenomena such as skyrmions, DWs, and non-topological droplets, etc. The continuous deformation between topologically inequivalent  skyrmions and  DWs may not only add
an additional degree of freedom but also introduce new functionalities for ultradense memory and spin logic applications.
A hybrid device (Fig. 5) comprised of skyrmions and DW pairs will have the combined functionalities of individual DW and skyrmion.
As illustrated in Fig. 5, a train of DWs can be generated by  applying a field or a current pulse.
The information (i.e., a series of bit) can be encoded as a train of DWs,
which will propagate along the nanowire driven by an external magnetic field or STT, and then will be converted into a train of skyrmions in the channel.
In this skyrmion channel, a functional manipulation of skyrmions is to be implemented, which is beyond the scope of the present work.
In the drain, skyrmions can be converted back to DWs, enabling read-out of the information encoded in the carrier through the giant-magnetoresistance (GMR) or tunneling magnetoresistance (TMR) effects.
A precise control of the time sequence of multi-skyrmions (or skyrmoin array) is possible by controlling the magnetic field to create a sequence of DW pairs.

\textbf{Writability} The nucleation of isolated skyrmions is a challenging task since it necessarily involves overcoming the topological barrier.
 In contrast, a variety of writing schemes have been demonstrated to generate a desired pattern of DWs by applying magnetic field and currents pulse \cite{Parkin,Hayashi}. In this regard, it is  efficient to encode the information in the DW sequence in the source (Fig. 5) since a DW based device has relatively better writability performance than skyrmions.

\textbf{Power consumption} One challenging task of spintronics community is to reduce the power consumption and improve the energy efficiency. The critical current density for driving the skyrmions (10$^{-7}$$-$10$^{-6}$ $\mu$A/nm$^2$), which is 4-5 orders smaller than the critical current density for moving DWs (10$^{-3}$ $-$ 1 $\mu$A/nm$^2$) \cite{IwasakiNcom}.

\textbf{Velocity and transmission}
The maximum velocity of DW (up to a few hundred meter per second) is larger than that of skyrmions (up to ~100 m/s), provided the applied current density exceeding the threshold values for both cases. However, the slope of current density vs. velocity (\emph{j}$-$\emph{v}) curve of skyrmions is much larger than that of DWs. Furthermore, it is shown that the pinning effect for skyrmions motion is much weaker, leading to better mobility and controllability of skyrmions than DWs. Therefore, it is more efficient to develop an information transmission channel based on skyrmions.

\textbf{Reliability}
The stability of skyrmions is protected topologically once they are created. Therefore it is superior to transmit information in the channels based on skyrmions as shown in Fig. 5.

\textbf{Readout}
The readout of both skyrmions and DWs can be realized by the GMR or TMR effect. However, only a small fraction of spins in the skyrmion core are reversed for generating MR signal. Therefore the readout of DW may be advantageous since it makes use of all the fully reversed spins inside the DW.

\section{Methods}

\subsection*{Micromagnetics simulations}
We have numerically solved the Landau-Lifshitz-Gilbert (LLG) equation, which governs the dynamics of the spin $\boldsymbol{S}_i$ at the lattice site $i$,
\begin{align}
\frac{d\boldsymbol{S}_{i}}{dt}&=-\gamma \boldsymbol{B}_{i}^{\text{eff}}\times \boldsymbol{S}_{i}+\frac{\alpha }{S}\boldsymbol{S}_{i}\times \frac{d\boldsymbol{S}_{i}}{dt}+\frac{pa^3}{2eS}(\boldsymbol{j(r)}\cdot\nabla)\boldsymbol{S}_{i}\notag\\
&-\frac{pa^3\beta}{2eS^2}[\boldsymbol{S}_{i}\times (\boldsymbol{j(r)}\cdot\nabla)\boldsymbol{S}_{i}],
\label{EqLLG}
\end{align}%
where $\gamma =\left\vert e\right\vert /m$ is the gyromagnetic ratio and $%
\alpha $ is the Gilbert-damping coefficient originating from spin
relaxation. The effective magnetic field $\boldsymbol{B}_{i}^{\text{eff}}$
acting on the spin $\boldsymbol{S}_{i}$ is determined by the Hamiltonian $H$ as%
\begin{equation}
\boldsymbol{B}_{i}^{\text{eff}}\equiv -\frac{1}{\hbar \gamma }\frac{\delta }{%
\delta \boldsymbol{S}_{i}}H.  \label{EffecB}
\end{equation}%
The third and fourth terms describe the coupling between the spin and the spin-polarized electric current $\boldsymbol{j(r)}$:
The third term describes the coupling via the spin transfer torque, while the fourth term via non-adiabatic effects.
The coefficient $\beta$ of the fourth term determines the strength of the nonadiabatic torque.
We have additionally included the dipole-dipole interactions (DDI) in all numerical calculations \cite{Newell}.
However the results remain quantitatively the same as those without the DDI.

\subsection*{Simulation parameters}
The nanowire is assumed to be a $1$~nm thick Cobalt on a substrate inducing DMI. The length of the nanowire is 1600 nm. While the width of the right half of the nanowire ($x\geqslant$ 800 nm) is fixed to be 100 nm, the left part ($x<$ 800 nm) is varied between 20 nm to 60 nm.  The following material parameters are adopted\cite{Sampaio}: exchange constant $A_{ex} = 15$~pJ/m, Gilbert damping $\alpha = 0.3$, saturation magnetisation $M_s = 580$ kA/m, $D = 3$~mJ/m$^2$ and $K_u = 0.7$~mJ/m$^3$. These parameters lead to a typical length scale $\lambda = A_{ex}/D = 5$~nm.  All of our simulations are performed with unit cell size 1 nm, which is sufficiently smaller than $\lambda$ to ensure the numerical accuracy. The nonadiabatic torque coefficient $\beta = 0.3$ is set to the same value as $\alpha$ such that the skyrmion moves along the central line of the nanotrack without additional transverse motion\cite{Sampaio}. All the simulations start with a uniform ferromagnetic state with all the spins along $+z$ direction due to the strong PMA.

\subsection*{Derrick-Hobart theorem}
The dynamical stability of a skyrmion is shown by the Derrick-Hobart theorem\cite{Derrick,Hobart} by examining the scaling property.
The Hamiltonian consists of the the nonlinear O(3) sigma model $H_X$ describing the exchange energy, the easy-axis anisotropy term $H_A$, the DMI term $H_{DM}$ and the Zeeman term $H_Z$.
Suppose there exists a soliton solution $\boldsymbol{n}^{0}(\boldsymbol{x}) $ to the system.
With the use of it, we calculate each contribution as $E_{X}^{0}$, $E_{A}^{0}$, $E_{DM}^{0}$ and $E_{Z}^{0}$.
Now we consider the configuration $\boldsymbol{n}(\boldsymbol{x})=\boldsymbol{n}_{0}(\lambda \boldsymbol{x})$.
Substituting this into each term, we find the energy of the new configuration to be
\begin{equation}
E(\lambda )=E_{X}^{0}+\lambda ^{-2}E_{A}^{0}-\lambda
^{-1}|E_{DM}^{0}|+\lambda ^{-2}E_{Z}^{0}.
\end{equation}%
This function has a unique minimum point at
\begin{equation}
\lambda =2(E_{A}^{0}+E_{Z}^{0})/|E_{DM}^{0}|,  \label{CondiLambda}
\end{equation}%
where $\lambda =1$ for consistency. The scale is uniquely fixed.
From this theorem it is clear that a soliton solution is stabilized by the DMI.
On the other hand, we find that $\lambda\rightarrow\infty$ as  $E_{DM}^{0}\rightarrow0$ from equation (\ref{CondiLambda}),
which implies that a soliton shrinks to zero without the DMI.

\section*{Acknowledgements}
Y.Z. thanks the support by the UGC Grant AoE/P-04/08 of Hong Kong SAR government. M. E. thanks the support by the Grants-in-Aid for Scientific Research from the Ministry of Education, Science, Sports and Culture No. 25400317. M. E. is very much grateful to N. Nagaosa for many helpful discussions on the subject.

\section*{Contributions}
Y.Z. performed the micromagnetics simulations. M. E. performed the theoretical analysis. Both authors co-analysed the simulation results and co-wrote the manuscript.

\section*{Competing Interests}
The authors declare that they have no competing financial interests.

\section*{Correspondence}
Correspondence and requests for materials should be addressed to Yan Zhou~(yanzhou@hku.hk) or Motohiko Ezawa (ezawa@ap.t.u-tokyo.ac.jp).

\end{document}